\documentclass[cameraready]{Interspeech}

\title{ProsoCodec: Prosody-Oriented Speech Codec for Voice Conversion}

\author[affiliation={1}]{Jeongsoo}{Choi}
\author[affiliation={2}]{Ji-Hoon}{Kim}
\author[affiliation={3}]{Shujie}{Hu}
\author[affiliation={1}, correspondingauthor]{Joon Son}{Chung}

\address{
    $^1$ KAIST, South Korea \\
    $^2$ Chung-Ang University, South Korea \\
    $^3$ The Chinese University of Hong Kong, China
}

\email{jeongsoo.choi@kaist.ac.kr, joonson@kaist.ac.kr}


\usepackage{comment}

\usepackage{hyperref}
\usepackage{url}

\usepackage{booktabs}       
\usepackage{amsfonts}       
\usepackage{nicefrac}       
\usepackage{microtype}      
\usepackage{xcolor}         
\usepackage{colortbl}
\usepackage{graphicx} 
\usepackage{amsmath}
\usepackage{adjustbox}
\usepackage{multirow}
\usepackage{enumitem}
\usepackage{makecell}
\usepackage{arydshln}

\begin{document}

\maketitle

\begin{abstract}
Neural speech codecs efficiently compress speech and have become a foundation for speech generation, but they are typically learned as holistic representations that intertwine linguistic content, speaker identity, and prosody. While this design is effective for zero-shot voice cloning, it hinders downstream tasks that require prosody preservation or transfer, such as voice conversion. To address this, we introduce ProsoCodec, a prosody-oriented speech codec that models prosody as a conditional residual rather than as a disentangled stream. Specifically, by conditioning both the encoder and decoder on text and speaker embeddings as prefix tokens, the discrete bottleneck is encouraged to capture prosodic variation not explained by content and speaker. To further preserve prosody, we use the low-frequency mel band and train the model on paired same-speaker utterances. Experiments on voice conversion show improved prosody preservation and reduced source-timbre leakage.
\end{abstract}

\section{Introduction}
Recent advancements in zero-shot speech generation and voice cloning have demonstrated remarkable capabilities in producing high-fidelity speech from brief reference prompts~\cite{chen2025neural, anastassiou2024seed, meng2025autoregressive, du2025cosyvoice}. Much of this progress has been driven by neural speech codecs~\cite{defossez2023high, ji2025wavtokenizer}, which discretize continuous speech into compact tokens, serving as the foundation for frameworks employing language models for in-context learning~\cite{chen2025neural, peng24voicecraft, ye2025llasa}. However, to faithfully reconstruct speech, these codecs are typically learned as holistic representations that intertwine linguistic content, speaker identity, and prosody. While effective for zero-shot voice cloning, which aims to replicate the reference entirely, this entanglement makes it difficult to control prosody without altering content or timbre~\cite{ju2024naturalspeech, zhang2025vevo}.

This issue becomes particularly evident in voice conversion, where the goal is to modify speaker timbre while strictly preserving the source utterance’s linguistic content and prosodic variation~\cite{wang2020one, zhang2025vevo}. Previous approaches attempt to achieve this by decomposing speech into distinct attributes such as content, timbre, and pitch~\cite{polyak2021speech, choi2021neural}, or by learning disentangled representations through restrictive bottlenecks~\cite{qian2020unsupervised} or adversarial objectives~\cite{ju2024naturalspeech} within an autoencoder. Voice conversion is then performed by replacing the speaker representation while keeping the remaining factors fixed. However, prosody is not a standalone attribute that can be cleanly swapped, as it is conditioned on both content and speaker, varying with context even within a single speaker~\cite{skerry2018towards}. Consequently, enforcing a fully speaker-independent prosody representation often discards speaker-conditioned prosodic nuances and limits expressiveness~\cite{qu2025disentanglement}.

To overcome this, we propose ProsoCodec, a prosody-oriented speech codec conditioned on text and speaker identity to encode residual prosodic variation. Unlike neural speech codecs that aim to faithfully reconstruct holistic acoustic signals under a limited bitrate~\cite{defossez2023high, li2024single, ji2025wavtokenizer, parker2025scaling}, our objective is selective representation learning tailored for fine-grained prosodic control. Leveraging the strong performance of recent automatic speech recognition (ASR)~\cite{radford2023robust, shi2026qwen3} and speaker verification (SV)~\cite{chen2022wavlm, wang2023cam++} systems, linguistic content and global speaker identity have become highly reliable and easily accessible features. By conditioning both the encoder and decoder on these attributes and introducing a discrete bottleneck through quantization, ProsoCodec encourages the latent representation to focus on residual prosodic variation rather than redundant content or speaker information. This conceptual shift allows the model to achieve effective residual modeling without relying on fragile disentanglement mechanisms such as adversarial objectives.

ProsoCodec builds upon a diffusion autoencoder architecture~\cite{wang2025tadicodec, wang2026scaling}, where an encoder maps speech into discrete tokens and a diffusion decoder reconstructs high-fidelity mel-spectrograms from them. We restrict the input mel-spectrogram to its low-frequency band to bias the tokens toward prosodic variation, inspired by previous works~\cite{ren2022prosospeech, jiang2023mega}. For voice conversion, the decoder generates speech from the source tokens while conditioning on a reference prompt to adopt the target speaker timbre. However, such prompt-based generation tends to imitate the reference holistically, and can therefore override the source prosody~\cite{pei2026unified}. To mitigate this, we introduce a dual-utterance training strategy using paired utterances from the same speaker, making utterance-level style harder to copy from the prompt, thereby improving prosody preservation. Extensive experiments demonstrate that ProsoCodec outperforms previous zero-shot voice conversion methods in prompt speaker similarity, source content preservation, and source prosody preservation. We further conduct ablation studies to provide insights into how different design components influence voice conversion performance.

Our contributions are three-fold: 1) We propose ProsoCodec, a novel generative speech codec that models prosody as a conditional residual by explicitly conditioning on text and speaker identity, without requiring hand-crafted prosody supervision or complex adversarial disentanglement. 2) We introduce a simple yet effective dual-utterance training strategy to explicitly prevent prompt-style leakage, decoupling utterance-level prosody from global speaker timbre. 3) We present comprehensive ablation studies on input representations and bottleneck designs, offering insights into balancing timbre conversion quality and source-style preservation for high-quality zero-shot voice conversion.

\begin{figure}[t]
  \centering
  \includegraphics[width=7cm]{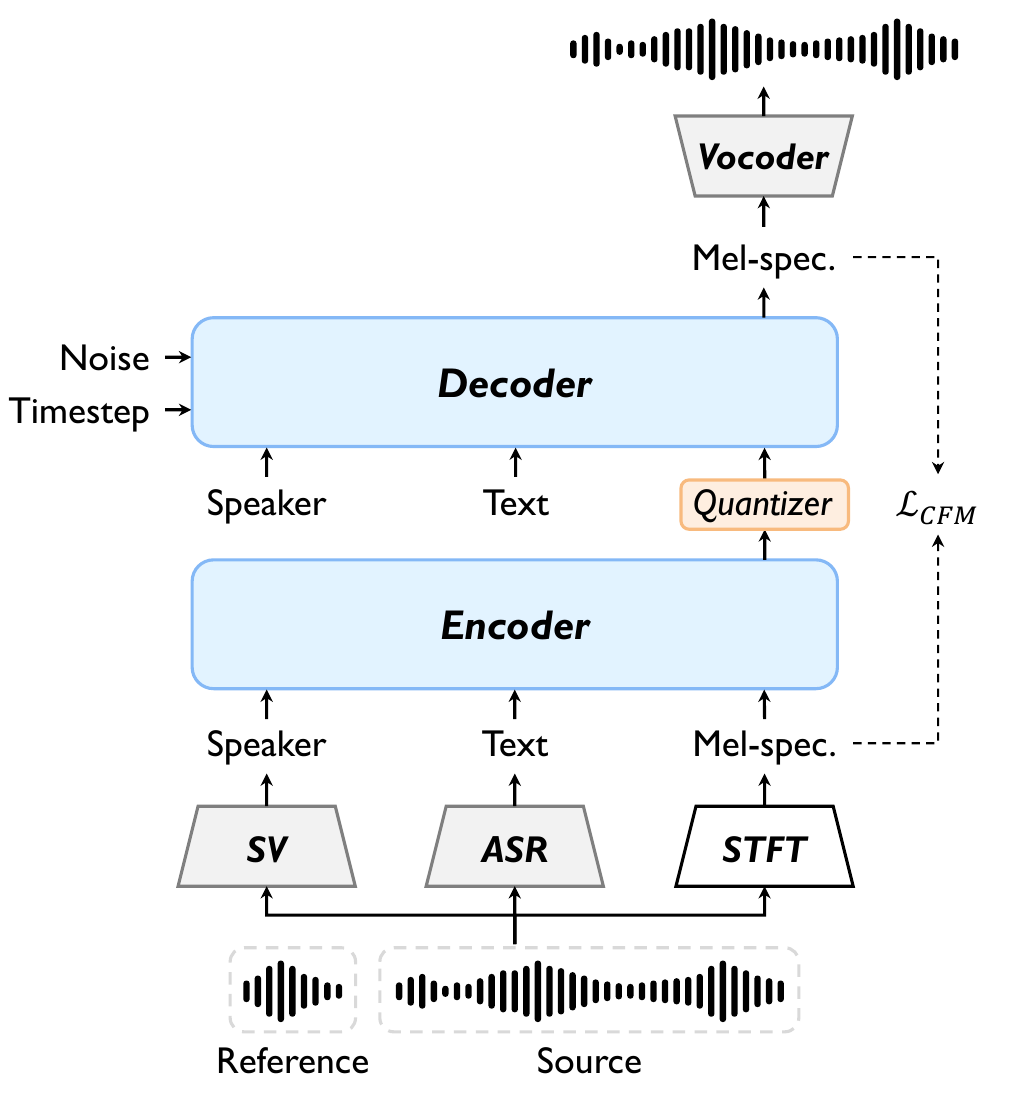}
  \vspace{-1mm}
  \caption{Model architecture of ProsoCodec. Reference and source are from the same speaker during training and different speakers during inference. The decoder receives a clean reference and a noisy source mel-spectrogram.}
  \vspace{-2mm}
  \label{fig:1}
\end{figure}

\section{Method}
\subsection{Overview}
Given a source utterance and a reference utterance, our voice conversion framework aims to generate speech that preserves the linguistic content and prosodic style of the source while adopting the speaker timbre of the reference. Both waveforms are separately encoded by the encoder of ProsoCodec and quantized into discrete token sequences. The decoder then generates the converted speech from the source tokens, which carry the source prosody, while the reference speech along with its tokens is provided as a prompt that anchors the target speaker timbre. To encourage the discrete tokens to retain prosodic style information rather than linguistic and timbre information, we additionally condition ProsoCodec on transcripts and speaker embeddings obtained from pretrained models. The overall architecture of ProsoCodec is illustrated in Fig.~\ref{fig:1}.

\subsection{ProsoCodec}
\subsubsection{Explicit Text and Speaker Priors}
To encourage the codec to capture residual information beyond linguistic and speaker content, we incorporate explicit text and speaker priors during both encoding and decoding. We first obtain utterance transcripts from an ASR model pretrained on large-scale, diverse data. The transcripts are tokenized into subwords, embedded via a lookup table, and processed by an MLP-based text encoder to match the module input space, producing $\mathbf{e}_{\text{txt}} \in \mathbb{R}^{L \times D}$, where $L$ is the token length and $D$ the model dimension. Speaker characteristics are extracted by a pretrained SV model as an utterance-level embedding and projected through an MLP-based speaker encoder to obtain $\mathbf{e}_{\text{spk}} \in \mathbb{R}^{1 \times D}$.

\subsubsection{Prior-guided Encoder}
The encoder compresses the input speech waveform, conditioned on the explicit priors. Given an input waveform, we extract a mel-spectrogram using the short-time Fourier transform (STFT) and project it to the encoder dimension through a linear layer, resulting in $\mathbf{e}_{\text{mel}} \in \mathbb{R}^{T \times D}$, where $T$ denotes the temporal length. The conditioning signals ($\mathbf{e}_{\text{spk}}$ and $\mathbf{e}_{\text{txt}}$) are utilized as prefix tokens and concatenated with the mel-spectrogram along the temporal dimension: $\mathbf{e} = [\mathbf{e}_{\text{spk}}; \mathbf{e}_{\text{txt}}; \mathbf{e}_{\text{mel}}]$, where $\mathbf{e} \in \mathbb{R}^{(1 + L + T) \times D}$. The concatenated sequence is processed by a Transformer~\cite{vaswani2017attention}-based encoder with bidirectional self-attention, allowing the mel-spectrogram to be encoded while being interpreted based on the prefix information. The guided mel features are obtained by slicing the encoder output:
\begin{equation}
    \mathbf{h} = \text{Encoder}(\mathbf{e}), \quad
    \mathbf{h}' = \mathbf{h}_{1+L:1+L+T}.
\end{equation}

\subsubsection{Quantizer}
While the explicit priors provide strong guidance regarding linguistic content and speaker identity, the continuous representation $\mathbf{h}'$ possesses sufficient capacity to retain entangled acoustic details~\cite{zhang2025vevo}. To encourage the model to encode the residual prosodic style, we introduce a strict information bottleneck via quantization. By restricting the representation capacity to a discrete space, this bottleneck discourages the model from allocating limited resources to redundant information. Instead, it forces the model to rely on the explicit priors, directing its discrete tokens to capture the structural variations corresponding to prosody, such as intonation and rhythm. To further control this information bottleneck via the token rate, we apply linear interpolation before and after quantization. We employ binary spherical quantization (BSQ)~\cite{zhao2025image}, which maps the latents to a discrete token sequence by projecting features onto an $\ell_2$-normalized hypersphere and performing binary quantization:
\begin{equation}
    \mathbf{h}_q = \text{Quantizer}(\text{Interp}(\mathbf{h}')).
\end{equation}
The quantized output $\mathbf{h}_q \in \{-1,1\}^{T' \times d}$ has sequence length $T'$ after interpolation and projected dimension $d$, producing an implicit binary codebook of size $2^d$.

\begin{table*}[t]
    \centering
    \caption{Objective and subjective results of different methods for voice conversion. The merged evaluation set contains LibriTTS test-clean, test-other, and VCTK. Best and second-best results are in \textbf{bold} and \underline{underlined}, respectively.}
    \vspace{-1mm}
    \label{table1}
    \setlength{\tabcolsep}{5pt}
    \resizebox{0.8\linewidth}{!}{
    \begin{tabular}{lcccccccc}
    \toprule
    & \multicolumn{1}{c}{\textit{Content}}
    & \multicolumn{3}{c}{\textit{Timbre}}
    & \multicolumn{2}{c}{\textit{Prosody}}
    & \multicolumn{2}{c}{\textit{Naturalness}} \\
    \cmidrule(lr){2-2} \cmidrule(lr){3-5} \cmidrule(lr){6-7} \cmidrule(lr){8-9}
    \textbf{Model} & \textbf{WER$\downarrow$} & \textbf{SIM$_\text{r}$$\uparrow$} & \textbf{SIM$_\text{s}$$\downarrow$} & \textbf{S-MOS$\uparrow$} & \textbf{RMSE$\downarrow$} & \textbf{P-MOS$\uparrow$} & \textbf{UTMOS$\uparrow$} & \textbf{N-MOS$\uparrow$} \\
    \midrule
    \rowcolor{gray!15}
    Source Speech & 3.668 & 0.090 & 1.000 & - & 0.000 & - & 3.897 & 4.253{\scriptsize~$\pm$~0.136} \\
    DDDM-VC~\cite{choi2024dddm} & 9.186 & 0.297 & 0.296 & 3.309{\scriptsize~$\pm$~0.139} & 0.472 & 3.012{\scriptsize~$\pm$~0.174} & 3.624 & 2.006{\scriptsize~$\pm$~0.152} \\
    UniAudio~\cite{yang2024uniaudio} & 9.076 & 0.276 & \underline{0.178} & 2.809{\scriptsize~$\pm$~0.175} & 0.470 & 2.475{\scriptsize~$\pm$~0.163} & 3.631 & 3.006{\scriptsize~$\pm$~0.174} \\
    HierSpeech++~\cite{lee2025hierspeech++} & 7.568 & 0.381 & 0.312 & 3.494{\scriptsize~$\pm$~0.148} & \underline{0.461} & 3.389{\scriptsize~$\pm$~0.170} & \textbf{3.869} & 2.679{\scriptsize~$\pm$~0.167} \\
    FACodec~\cite{ju2024naturalspeech} & 5.454 & 0.354 & 0.347 & 3.247{\scriptsize~$\pm$~0.169} & 0.455 & 3.364{\scriptsize~$\pm$~0.193} & 3.544 & 2.302{\scriptsize~$\pm$~0.172} \\
    Seed-VC~\cite{liu2024zero} & 5.078 & \underline{0.531} & 0.239 & \textbf{4.210}{\scriptsize~$\pm$~0.124} & 0.473 & 3.463{\scriptsize~$\pm$~0.156} & 3.712 & 3.611{\scriptsize~$\pm$~0.163} \\
    Vevo~\cite{zhang2025vevo} & \underline{4.826} & 0.478 & 0.243 & 4.130{\scriptsize~$\pm$~0.139} & 0.464 & \underline{3.506}{\scriptsize~$\pm$~0.155} & 3.800 & \textbf{4.056}{\scriptsize~$\pm$~0.144} \\
    \rowcolor{blue!10}
    ProsoCodec (Ours) & \textbf{4.451} & \textbf{0.565} & \textbf{0.167} & \underline{4.167}{\scriptsize~$\pm$~0.134} & \textbf{0.428} & \textbf{3.852}{\scriptsize~$\pm$~0.172} & \underline{3.853} & \underline{4.000}{\scriptsize~$\pm$~0.148} \\
    \bottomrule
    \end{tabular}
    }
    \vspace{-1mm}
\end{table*}

\subsubsection{Diffusion-based Decoder}
To reconstruct high-fidelity mel-spectrograms from the prosody-oriented discrete tokens, we adopt a diffusion-based decoder. The decoder is based on Diffusion Transformer (DiT)~\cite{peebles2023scalable}, which shares a similar architecture with the encoder but additionally incorporates a diffusion timestep. The timestep embedding is injected into every layer via adaptive layer normalization, indicating the current noise level. Given a clean mel-spectrogram $\mathbf{x} \in \mathbb{R}^{T \times 128}$, we sample Gaussian noise $\boldsymbol{\epsilon}$ and construct a noisy input by interpolation, $\mathbf{x}_t = t\,\mathbf{x} + (1-t)\boldsymbol{\epsilon}
$, where $t \sim \mathcal{U}(0,1)$. Following the speech infilling objective of recent flow-matching models~\cite{le2023voicebox, chen2025f5}, we apply a random span mask $\mathbf{m} \in \{0,1\}^{T}$ to simulate the inference setting where reference speech is provided as a prompt. The masked noisy mel-spectrogram is defined as
\begin{equation}
    \tilde{\mathbf{x}}_t = \mathbf{m} \odot \mathbf{x}_t + (1-\mathbf{m}) \odot \mathbf{x},
\end{equation}
where $\odot$ denotes element-wise multiplication. The noisy mel-spectrogram is projected to the model dimension as $\mathbf{e}_{noisy} \in \mathbb{R}^{T \times D}$. The quantized representations $\mathbf{h}_q$ are projected and temporally interpolated to match the same length, yielding $\mathbf{e}_{q} \in \mathbb{R}^{T \times D}$. Similar to the encoder, the decoder receives the concatenated sequence $\mathbf{e} = [\mathbf{e}_{\text{spk}}; \mathbf{e}_{\text{txt}}; \mathbf{e}_{\text{noisy}} + \mathbf{e}_{q}]$,
\begin{equation}
    \mathbf{h} = \text{Decoder}(\mathbf{e}), \quad
    \mathbf{h}' = \mathbf{h}_{1+L:1+L+T}.
\end{equation}
Finally, $\mathbf{h}'$ is linearly projected to predict the diffusion velocity field. We optimize the decoder with a flow-matching objective, encouraging the predicted velocity to match the target $(\mathbf{x}-\boldsymbol{\epsilon})$.

\subsection{Training and Inference}
The encoder, quantizer, and decoder of our model are trained in an end-to-end manner with conditional flow matching loss:
\begin{equation}
    \mathcal{L}_{CFM} = \mathbb{E}_{t}||v_t(\mathbf{x}_t|\mathbf{e}, \theta) - u_t(\mathbf{x}_t)||^2,
\end{equation}
where $\theta$ refers to the DiT decoder, $v_t$ represents the estimated vector fields, and $u_t$ is the target vector field.

During inference for voice conversion, both the source and reference speech are separately encoded into discrete token sequences. The decoder then generates the converted speech conditioned on the reference speaker embedding, the source transcript, and the source tokens. Here, the reference speech, together with its transcript and tokens, is provided as a prompt to facilitate in-context consistency.

\subsection{Training Strategies for Prosody Preservation}
Directly extending random span masking and inpainting to voice conversion can cause prompt-style leakage. Since the prompt and source come from the same utterance, and prompt-based generation holistically imitates the prompt~\cite{pei2026unified}, its style leaks into the output rather than the source's prosody being preserved. To mitigate this issue, we introduce a dual-utterance training strategy that samples paired utterances from the same speaker, using one as the prompt and the other as the source. This leads to better preservation of the source prosody at inference. We alternate this strategy with random span masking during training.

Furthermore, since we do not impose an explicit disentanglement objective, the bottleneck risks retaining excessive acoustic detail to minimize the flow-matching loss. As prosodic cues primarily reside in the low-frequency region of the mel-spectrogram~\cite{ren2022prosospeech, jiang2023mega}, we feed only that band to the encoder, biasing the tokens toward prosody. In contrast, the prompt fed to the decoder uses the full-band mel-spectrogram to provide speaker characteristics.

\section{Experiments}
\subsection{Datasets}
Our model is trained on the LibriTTS~\cite{zen2019libritts} dataset, which comprises 585 hours of 24\,kHz read speech from 2,456 speakers. To evaluate its performance, we utilize the LibriTTS \texttt{test-clean} and \texttt{test-other} splits, alongside the VCTK dataset~\cite{yamagishi2019cstr} for assessing out-of-domain generalization. From each evaluation set, we randomly sample 1,000 utterances with durations between 2 and 8 seconds. For each sample, we select a reference prompt from the same speaker for speech resynthesis, and another from a different speaker for zero-shot voice conversion.

\subsection{Implementation Details}
The speech waveform is converted into a mel-spectrogram using 128 mel filter banks with an 80\,ms window length and 20\,ms hop size, resulting in a frame rate of 50\,Hz. We use pretrained Qwen3-ASR~\cite{shi2026qwen3} and CAM++~\cite{wang2023cam++} to obtain transcripts and speaker embeddings, respectively. ProsoCodec comprises a Transformer encoder~\cite{vaswani2017attention} and a Diffusion Transformer (DiT) decoder~\cite{peebles2023scalable}, both initialized from TaDiCodec~\cite{wang2025tadicodec} for efficient training. The encoder and decoder consist of 8 and 16 layers, respectively, each with a hidden size of 1024, an intermediate size of 4096, and 16 attention heads. We train the model with the AdamW~\cite{loshchilov2019decoupled} optimizer using a learning rate of $2\times10^{-5}$, $\beta_1=0.9$, $\beta_2=0.999$, and a weight decay of $0.01$. The learning rate is linearly warmed up over the first 10k updates, and training is conducted for 150k updates with a global batch size of 160 seconds. During inference, mel-spectrograms are generated using 32 sampling steps via the Euler ODE solver and subsequently converted into 24\,kHz waveforms by a Vocos vocoder~\cite{siuzdak2024vocos}.

\subsection{Evaluation Metrics}
The quality of the synthesized speech is assessed through a comprehensive suite of subjective and objective metrics. Given that subjective evaluation remains the gold standard for synthesized speech~\cite{saeki2024speechbertscore}, we conducted Mean Opinion Score (MOS) tests. For each test subset, 15 listeners rated 10 randomly sampled utterances on a 5-point Likert scale. We specifically measured three criteria: speaker similarity (\textbf{S-MOS}) to the reference, prosody similarity (\textbf{P-MOS}) to the source, and overall naturalness (\textbf{N-MOS}). For objective evaluation, we compute Word Error Rate (\textbf{WER}) by comparing the source transcript with the automatic transcript of the generated sample. To avoid evaluation bias, we utilize Whisper-large-v3~\cite{radford2023robust}, which is different from the ASR model used in ProsoCodec. To evaluate speaker similarity, we compute the cosine similarity between the speaker embeddings of the reference and synthesized speech ($\textbf{SIM}_\textbf{r}$), where the embeddings are extracted using the pre-trained WavLM-Large model \cite{chen2022wavlm}. Furthermore, to quantify potential speaker identity leakage during voice conversion, we report the $\textbf{SIM}_\textbf{s}$ between the source and converted speech; here, lower similarity scores indicate superior quality. For prosody preservation, we adopt fundamental frequency ($f_0$) as a proxy for prosody dynamics. Specifically, we compute the Root Mean Square Error (\textbf{RMSE}) between the source and the generated speech, where $f_0$ values are extracted via Harvest~\cite{morise2017high} and log-scaled to align with human perception. \textbf{UTMOS}~\cite{saeki2022utmos} is used to estimate perceptual naturalness of the generated speech, serving as an automatic MOS predictor.

\begin{table}[t]
    \centering
    \caption{Ablation study about zero-shot voice conversion on LibriTTS test-clean dataset.}
    \vspace{-1mm}
    \label{table2}
    \setlength{\tabcolsep}{4pt}
    \resizebox{\linewidth}{!}{
    \begin{tabular}{lccccc}
        \toprule
        \textbf{Method} & \textbf{WER$\downarrow$} & \textbf{SIM$_\text{r}$$\uparrow$} & \textbf{SIM$_\text{s}$$\downarrow$} & \textbf{RMSE$\downarrow$} & \textbf{UTMOS$\uparrow$} \\
        \midrule
        \rowcolor{gray!15}
        Source Speech                & 3.921  & 0.081 & 1.000 & 0.000 & 4.091 \\
        \rowcolor{blue!10}
        ProsoCodec                   & 4.684  & 0.608 & 0.168 & 0.376 & 3.978 \\
        ~~$-$ low-freq. mel          & 4.810  & 0.585 & 0.192 & 0.386 & 4.004 \\
        ~~~~~~$-$ dual-utter.        & 5.630  & 0.634 & 0.166 & 0.421 & 3.884 \\
        ~~~~~~~~~~$-$ speaker cond.  & 6.793  & 0.600 & 0.186 & 0.397 & 3.891 \\
        ~~~~~~~~~~~~~~$-$ text cond. & 86.601 & 0.569 & 0.108 & 0.434 & 3.569 \\
        \bottomrule
    \end{tabular}
    \vspace{-1mm}
    }
\end{table}

\begin{table}[t]
    \centering
    \caption{Ablation study about bottleneck configuration on LibriTTS test-clean dataset. FR: frame rate in Hz; CS: codebook size; BR: bitrate in bps.}
    \vspace{-1mm}
    \label{table3}
    \setlength{\tabcolsep}{2pt}
    \resizebox{\linewidth}{!}{
    \begin{tabular}{ccc ccc cccc}
        \toprule
        & & & \multicolumn{3}{c}{\textbf{Resynthesis}} & \multicolumn{4}{c}{\textbf{Voice Conversion}} \\
        \cmidrule(lr){4-6} \cmidrule(lr){7-10}
        \textbf{FR} & \textbf{CS} & \textbf{BR}
        & \textbf{WER$\downarrow$} & \textbf{SIM$\uparrow$} & \textbf{RMSE$\downarrow$}
        & \textbf{WER$\downarrow$} & \textbf{SIM$_\text{r}$$\uparrow$} & \textbf{SIM$_\text{s}$$\downarrow$} & \textbf{RMSE$\downarrow$} \\
        \midrule
        25   & 4096  & 300 & 4.28 & 0.72 & 0.21 & 4.49 & 0.56 & 0.20 & 0.36 \\
        12.5 & 16384 & 175 & 4.54 & 0.71 & 0.22 & 4.71 & 0.60 & 0.18 & 0.37 \\
        \rowcolor{blue!10}
        12.5 & 4096  & 150 & 4.30 & 0.71 & 0.22 & 4.68 & 0.61 & 0.17 & 0.38 \\
        12.5 & 1024  & 125 & 4.57 & 0.70 & 0.22 & 4.91 & 0.62 & 0.16 & 0.38 \\
        6.25 & 4096  & 75  & 4.85 & 0.69 & 0.23 & 6.17 & 0.64 & 0.14 & 0.41 \\
        \hdashline
        \multicolumn{3}{c}{\textit{Zero-Shot TTS}} & 5.85 & 0.67 & 0.29 & 6.04 & 0.68 & 0.11 & 0.48 \\
        \bottomrule
    \end{tabular}
    }
    \vspace{-1mm}
\end{table}

\section{Results}
\subsection{Comparisons with Previous Methods}
We compare ProsoCodec with several state-of-the-art open-source voice conversion models. In particular, we employ DDDM-VC~\cite{choi2024dddm}, HierSpeech++~\cite{lee2025hierspeech++}, and Vevo~\cite{zhang2025vevo}, which are resynthesis-based models that utilize decomposed speech attributes; UniAudio~\cite{yang2024uniaudio}, an LLM-based unified audio generation framework built on a neural audio codec; FACodec~\cite{ju2024naturalspeech}, a factorized speech codec-based model; and Seed-VC~\cite{liu2024zero}, a prompt-based diffusion model trained with a timbre-shift strategy. Table~\ref{table1} reports the objective and subjective results on the merged evaluation set. The proposed ProsoCodec model demonstrates superior performance in terms of WER, indicating that the linguistic content of the source speech is well preserved during conversion. The lower SIM$_{\text{s}}$ and higher SIM$_{\text{r}}$ further show that our method effectively suppresses timbre leakage from the source speaker while adopting the target speaker timbre.

These improvements can be attributed to several design choices in ProsoCodec, including explicit conditioning on robust transcripts and speaker embeddings, the prompt-based conversion mechanism with dual-utterance training strategy, and the use of low-frequency band mel-spectrograms. We analyze the contribution of each component in the ablation studies.

\subsection{Ablation Study}
Table~\ref{table2} presents a cumulative ablation on the LibriTTS test-clean dataset to analyze the contribution of each component in ProsoCodec. When the low-frequency mel-spectrogram is replaced with full-band mel-spectrogram input, the model shows degraded performance across several metrics, suggesting that restricting the acoustic bandwidth helps the codec focus on prosodic variations rather than fine-grained spectral details. Removing the dual-utterance training strategy further degrades prosody preservation, as reflected by the increased RMSE. This result suggests that the strategy encourages the codec to retain prosodic information required for generation by preventing the model from trivially copying information from the prompt. When speaker conditioning is removed, speaker similarity drops, reflected by decreased SIM$_\text{r}$ and increased SIM$_\text{s}$, indicating stronger leakage from the source speaker. Finally, removing text conditioning leads to a substantial increase in WER. These results highlight that explicit text and speaker conditioning play critical roles in guiding the codec to encode residual information within the limited bitrate of the discrete bottleneck.

\begin{figure}[t]
  \centering
  \includegraphics[width=\columnwidth]{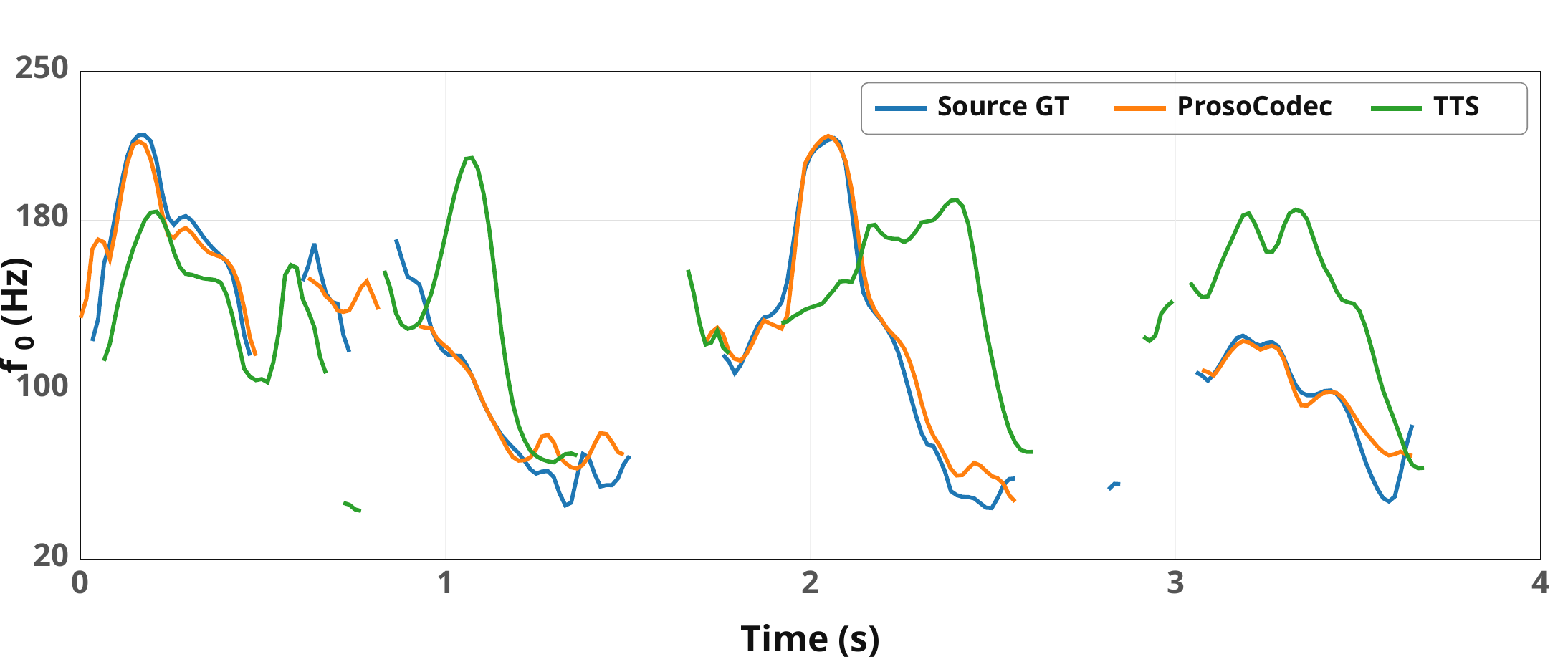}
  \vspace{-5mm}
  \caption{Pitch contour of resynthesis result of ProsoCodec along with source speech and zero-shot TTS output.}
  \vspace{-1mm}
  \label{fig:2}
\end{figure}

\subsection{Analysis}
Table~\ref{table3} presents the ablation study on the codec bottleneck configuration. Consistent with observations in previous studies, strong reconstruction does not necessarily imply strong conversion. Increasing the bitrate improves resynthesis performance as more acoustic details can be preserved, however, this also leads to source-timbre leakage that harms voice conversion performance. Considering this trade-off, we select a bottleneck configuration of 12.5\,Hz frame rate with a codebook size of 4096 (150\,bps), which provides a good balance between content preservation and voice conversion quality.

We further analyze an extreme scenario where the decoder is trained entirely without codec tokens. Since it remains conditioned on the acoustic prompt, speaker embedding, and text, this setup effectively operates as a zero-shot TTS or voice cloning model~\cite{le2023voicebox, chen2025f5}. While this baseline successfully generates speech that follows the prompt's speaker characteristics well, it derives its prosodic patterns directly from the prompt or synthesizes generalized text-driven prosody. This behavior is also reflected in the pitch contours in Fig.~\ref{fig:2}. ProsoCodec closely resynthesizes the source $f_0$ trajectory, whereas the model without codec tokens produces pitch that deviates from the source. Such behavior is fundamentally misaligned with the objective of voice conversion, which requires mapping the exact prosodic variations of the source.

\section{Conclusion}
In this paper, we presented ProsoCodec, a prosody-oriented speech codec designed for high-fidelity zero-shot voice conversion. Explicitly conditioned on linguistic content and speaker identity, our model effectively captures residual prosodic variations without compromising speaker-conditioned nuances and expressiveness. We further introduced a dual-utterance training strategy and low-frequency input biasing to significantly mitigate the common issue of prompt-style leakage. Extensive evaluations demonstrate that ProsoCodec excels in preserving both source prosody and target speaker timbre, outperforming existing state-of-the-art frameworks. We believe that our approach to residual prosody modeling provides a robust foundation for prosodic representations, which can be widely utilized across various speech tasks, including expressive text-to-speech synthesis, voice conversion, and spoken dialogue systems.

\clearpage
\section{Acknowledgments}
This work was supported by Institute of Information \& communications Technology Planning \& Evaluation (IITP) grant funded by the Korean government (MSIT, RS-2025-02263977, Development of Communication Platform supporting User Anonymization and Finger Spelling-Based Input Interface for Protecting the Privacy of Deaf Individuals).

\section{Generative AI Use Disclosure}
Generative AI tools were used only for editing and polishing this manuscript and were not used for producing any significant part of the manuscript.

\bibliographystyle{IEEEtran}
\bibliography{mybib}

\end{document}